\documentclass[aps,preprintnumbers,eqsecnum,amsmath,amssymb,showpacs,nofootinbib]{revtex4}
\usepackage{graphicx}
\usepackage{dcolumn}
\usepackage{bm}
\usepackage{epsfig}

\begin{document} 
\title{The {\boldmath $\gamma^*\,\gamma^*\to\eta_c$} transition form factor}
\author{Wolfgang Lucha$^{1}$ and Dmitri Melikhov$^{1,2,3}$}
\affiliation{
$^1$HEPHY, Austrian Academy of Sciences, Nikolsdorfergasse 18, A-1050 Vienna, Austria\\
$^2$Faculty of Physics, University of Vienna, Boltzmanngasse 5, A-1090 Vienna, Austria\\
$^3$SINP, Moscow State University, 119991 Moscow, Russia}
\date{\today}

\begin{abstract}
We study the $\gamma^*\,\gamma^*\to\eta_c$ transition form factor,
$F_{\eta_c\gamma\gamma}(Q_1^2,Q_2^2),$ with the local-duality (LD)
version of QCD sum rules. We analyse the extraction of this
quantity from two different correlators, $\langle PVV\rangle$ and
$\langle AVV\rangle,$ with $P,$ $A,$ and $V$ being the
pseudoscalar, axial-vector, and vector currents, respectively. The
QCD factorization theorem for
$F_{\eta_c\gamma\gamma}(Q_1^2,Q_2^2)$ allows us to fix the
effective continuum thresholds for the $\langle PVV\rangle$ and
$\langle AVV\rangle$ correlators at large values of $Q^2=Q_2^2$
and some fixed value of $\beta\equiv Q_1^2/Q_2^2.$ We give
arguments that, in the region $Q^2\ge10$--$15\;\mbox{GeV}^2,$ the
effective threshold should be close~to~its asymptotic value such
that the LD sum rule provides reliable predictions for $F_{\eta_c\gamma\gamma}(Q_1^2,Q_2^2).$ 
We show that, for the experimentally relevant kinematics of one real and one virtual
photon, the result~of~the LD sum rule for $F_{\eta_c\gamma}(Q^2)\equiv F_{\eta_c\gamma\gamma}(0,Q^2)$ 
may be well approximated by the simple monopole formula $F_{\eta_c\gamma}(Q^2)={2e_c^2N_cf_P}(M_V^2+Q^2)^{-1},$ 
where $f_P$ is the $\eta_c$ decay constant, $e^2_c$ is the $c$-quark
charge, and the parameter $M_V$ lies in the mass range of
the~lowest $\bar cc$ vector states.
\end{abstract}
\maketitle

\section{Introduction}
The processes $\gamma^*\,\gamma^*\to P,$ with
$P=\pi^0,\eta,\eta',\eta_c,$ are of great interest for our
understanding of QCD and of the~meson structure. The corresponding
amplitude
\begin{eqnarray}
\langle\gamma^*(q_1)\gamma^*(q_2)|P(p)\rangle&=&{\rm
i}\epsilon_{\varepsilon_1\varepsilon_2 q_1
q_2}F_{P\gamma\gamma}(q_1^2,q_2^2)
\end{eqnarray}
contains only one invariant form factor,
$F_{P\gamma\gamma}(q_1^2,q_2^2),$ which is one of the simplest
hadronic form factors in QCD. We shall address the general
situation when both photons are virtual: $q_i^2=-Q_i^2,$
$Q_i^2\ge0,$ $i=1,2.$ From the experimental perspective, the most
interesting kinematical configuration is when one of the photons
is almost real and the~other has virtuality $Q^2.$ For this
special case, we use the notation $F_{P\gamma}(Q^2)\equiv
F_{P\gamma\gamma}(Q_1^2=0,Q_2^2=Q^2).$ The form factor
$F_{P\gamma}(Q^2)$ has been the subject of detailed experimental
\cite{cello,cleo,babar2006,babar2009,babar2010,babar2011,belle2012} and
theoretical investigations (for recent references, see
\cite{roberts,dorokhov,agaev,teryaev2,bt,kroll,mikhailov,blm,lcqm}).
A QCD factorization theorem predicts the behaviour of the form
factor at asymptotically large momentum~transfers~\cite{bl}:
\begin{eqnarray}
\label{factorization}
F_{P\gamma\gamma}(Q_1^2,Q_2^2)=2e_c^2\int\limits_0^1\frac{{\rm
d}\xi\phi^{\rm ass}_P(\xi)}{Q_1^2\xi+Q_2^2(1-\xi)},\qquad
\phi^{\rm ass}_P(\xi)=6f_P\xi(1-\xi),
\end{eqnarray}
which gives for $Q^2\equiv Q_2^2,$ $\beta\equiv Q_1^2/Q_2^2,$ and
$0\le\beta\le1$ (w.l.o.g, we denote the larger virtuality by
$Q_2^2$):
\begin{eqnarray}
\label{factorization2}
F_{P\gamma\gamma}(Q_1^2,Q_2^2)=\frac{6e_c^2f_P}{Q^2}\,I(\beta),\qquad
I(\beta)=\frac{1+2\beta\log\beta-\beta^2}{(1-\beta)^3},\qquad
I(0)=1,\qquad I(1)=1/3.
\end{eqnarray}
In the pion case, setting $Q_1^2=0$ and $Q_2^2=Q^2,$ this result
reduces to the asymptotic behaviour
$Q^2F_{\pi\gamma}(Q^2)\to\sqrt{2}f_\pi$ \cite{bl}, with
$f_\pi=0.130\;\mbox{GeV}.$ Similar relations follow for the mesons
$\eta$ and $\eta'$ after taking particle mixing into
account~\cite{anisovich,feldmann}.

Within errors, this saturation property is indeed found for the $\eta$ and $\eta'$ form factors. 
However, large-$Q^2$ data up to $Q^2=35\;\mbox{GeV}^2$ from {\sc BaBar} \cite{babar2009}
indicate that $Q^2F_{\pi\gamma}(Q^2)$ does not saturate at large $Q^2$ but increases further. 
No compelling theoretical explanation of the qualitatively different behaviour of the
$\pi\gamma$ form factor compared to the $\eta\gamma$ and $\eta'\gamma$ form factors has been proposed. 
As concluded in \cite{roberts,bt,mikhailov,blm}, the behaviour of the $\pi\gamma$
form factor is hard to explain in QCD. Moreover, the {\sc BaBar} findings for $F_{\pi\gamma}$ require
$O(1/s)$ duality-violating corrections between the hadron and the~QCD~spectral~densities~\cite{ms2012}.
Very recently, Belle \cite{belle2012} presented their results on the $F_{\pi\gamma}$ form factor 
which are in fact compatible with QCD factorization. 

Another, particularly interesting process is the transition $\gamma^*\,\gamma^*\to\eta_c.$ 
Here, one expects that, for the case of massive quarks, the onset of the factorization regime is,
compared to the case of massless quarks, delayed to higher $Q^2.$ The details of the form-factor behaviour provide valuable
information on the interplay of perturbative and nonperturbative QCD at intermediate and large momentum transfers.

In recent publications \cite{blm}, we analyzed the $P\gamma$ form
factors for light mesons, making use of QCD sum rules in their
local-duality (LD) limit \cite{ld}. We have given arguments that
the LD sum rules provide already for $Q^2$ larger~than~a~few
GeV$^2,$ reliable predictions for the $F_{P\gamma}$ form factors
of light pseudoscalars with an accuracy increasing very fast
with~$Q^2.$

The goal of this analysis is two-fold: First, we discuss the
subtleties of the formulation of a LD model for transition form
factors for the case of massive quarks. Second, we apply our LD
model to the case of the $\gamma^*\,\gamma^*\to\eta_c$ form
factor. The paper is organized as follows: Section~\ref{Sec:2PF}
briefly recalls results for the various 2-point functions that may
be used for the extraction of decay constants of heavy $\bar cc$
pseudoscalars. In Section~\ref{Sec:TFA}, we present the dispersion
representations for $\langle PVV\rangle$ and $\langle AVV\rangle$
3-point functions and discuss the procedure of obtaining the
$F_{P\gamma\gamma}(Q_1^2,Q_2^2)$ form factor from these
correlators. We also give our predictions for $F_{P\gamma}(Q^2)$
in a broad range of $Q^2.$ Section \ref{Sec:SAC} summarizes our
conclusions.

\section{Two-point functions of axial and pseudoscalar
currents}\label{Sec:2PF}For the case of massive quarks, one may
consider, on an equal footing, the $\langle AA\rangle,$ $\langle
AP\rangle,$ and $\langle PP\rangle$ correlators,~where
$A_\mu\equiv\bar\psi\gamma_\mu\gamma_5\psi$ and $P\equiv{\rm
i}\bar\psi\gamma_5\psi$ denote the axial-vector and pseudoscalar
currents, respectively.

The $\langle AA\rangle$ correlator involves two independent
Lorentz structures; to leading order in the strong coupling
$\alpha_s$ it~reads
\begin{eqnarray}
\left(p^2g_{\mu\nu}-p_\mu
p_\nu\right)\frac{N_c}{16\pi^2}\int\limits_{4m^2}^\infty\frac{{\rm
d}s}{s-p^2}\,\frac{4}{3}\left(1-\frac{4m^2}{s}\right)^{\!3/2}+p_\mu
p_\nu\,\frac{N_c}{16\pi^2}\int\limits_{4m^2}^\infty\frac{{\rm
d}s}{s-p^2}\,\frac{8m^2}{s}\sqrt{1-\frac{4m^2}{s}}.
\end{eqnarray}
We consider the sum rule for the longitudinal part $\langle
AA\rangle_L$ of the correlator $\langle AA\rangle,$ which contains
the contribution of the pseudoscalar mesons on its hadronic side.
The dispersion representations for the other correlators are
also~well-known \cite{svz,m}. After application of the Borel
transformation, one finds
\begin{align}
&\langle AA\rangle_L{:}\qquad f_P^2e^{-M^2\tau}+\mbox{excited
states}=\frac{N_cm^2}{2\pi^2}\int\limits_{4m^2}^\infty\frac{{\rm
d}s}{s}\,e^{-s\tau}\sqrt{1-\frac{4m^2}{s}}\left[1+O(\alpha_s)\right]
+\mbox{power corrections},
\\
&\langle AP\rangle{:}\qquad f_P^2M^2e^{-M^2\tau}+\mbox{excited
states}=\frac{N_cm^2}{2\pi^2}\int\limits_{4m^2}^\infty{\rm
d}s\,e^{-s\tau}\sqrt{1-\frac{4m^2}{s}}\left[1+O(\alpha_s)\right]
+\mbox{power corrections},
\\
&\langle PP\rangle{:}\qquad f_P^2M^4e^{-M^2\tau}+\mbox{excited
states}=\frac{N_cm^2}{2\pi^2}\int\limits_{4m^2}^\infty{\rm
d}s\,s\,e^{-s\tau}\sqrt{1-\frac{4m^2}{s}}\left[1+O(\alpha_s)\right]
+\mbox{power corrections}.
\end{align}
The sum rules for $\langle AP\rangle$ and $\langle PP\rangle$ may
be obtained from the $\langle AA\rangle$ sum rule by taking the
first and second $\tau$-derivatives, respectively. Thus,
considering any of these correlators leads to equivalent results
for the case of massive quarks, once proper subtractions are
performed.

Implementing quark--hadron duality in the usual way, {\em i.e.},
as a low-energy cut on the perturbative contribution to the
correlator, and setting $\tau=0$ (LD limit) --- in which case all
nonperturbative power corrections vanish ---,~the~resulting
expressions for the decay constants take the form
\begin{align}
\label{AA}&\langle AA\rangle_L{:}\qquad
f_P^2=\frac{N_cm^2}{2\pi^2}\int\limits_{4m^2}^{s_{\rm eff}^{\rm
AA}}\frac{{\rm d}s}{s}\sqrt{1-\frac{4m^2}{s}},
\\
\label{AP}&\langle AP\rangle{:}\qquad
f_P^2=\frac{N_cm^2}{2\pi^2}\int\limits_{4m^2}^{s_{\rm eff}^{\rm
AP}}\frac{{\rm d}s}{M^2}\sqrt{1-\frac{4m^2}{s}},
\\
\label{PP}&\langle PP\rangle{:}\qquad
f_P^2=\frac{N_cm^2}{2\pi^2}\int\limits_{4m^2}^{s_{\rm eff}^{\rm
PP}}{\rm d}s\,\frac{s}{M^4}\sqrt{1-\frac{4m^2}{s}}.
\end{align}
Obviously, the effective thresholds $s_{\rm eff}^{\rm AA},$
$s_{\rm eff}^{\rm AP},$ and $s_{\rm eff}^{\rm PP}$ must be
(slightly) different from each other.

It will be useful to recall that in the chiral limit, $m=0,$ the
situation is qualitatively different from the massive-quark case:
In the chiral limit, the $\langle AA\rangle$ correlator is
transverse and contains only one Lorentz structure,
$g_{\mu\nu}-p_\mu p_\nu/p^2.$ The corresponding invariant
amplitude contains the contribution of the Goldstone whereas
excited pseudoscalars decouple from the axial current in the
chiral limit \cite{svz}. Unlike the case of massless quarks,
massive ground-state~pseudoscalars do not contribute to the
transverse Lorentz structure of the $\langle AA\rangle$ correlator
of the axial currents of massive quarks~\cite{lm}.

\section{LD model for the $\gamma^*(Q_1)\,\gamma^*(Q_2)\to P$
transition form factor}\label{Sec:TFA}The $\gamma^*\,\gamma^*\to
P$ form factor may be extracted from two different correlators:
namely, from $\langle PVV\rangle$ and from $\langle AVV\rangle.$

\subsection{Transition form factor from the three-point function
$\langle PVV\rangle$}Let us start with the amplitude for
two-photon production from the vacuum $|0\rangle,$ induced by the
pseudoscalar~current $j^5(x)={\rm i}\bar\psi(x)\gamma_5\psi(x),$
with $\varepsilon_{1,2}$ denoting the photon polarization vectors:
\begin{equation}
\langle\gamma(q_1)\gamma(q_2)|j^5(x=0)|0\rangle=
T_{\alpha\beta}(p|q_1,q_2)\,\varepsilon^\alpha_1\varepsilon^\beta_2,
\qquad p\equiv q_1+q_2.
\end{equation}
The amplitude $T_{\alpha\beta}$ is obtained from the vacuum
expectation value of the $T$-product of one pseudoscalar and
two~vector currents and will be called the $\langle PVV\rangle$
amplitude. The decomposition of the amplitude contains only one
invariant~form factor $F_5$:
\begin{eqnarray}
\label{amp1}
T_{\alpha\beta}(p|q_1,q_2)&=&\epsilon_{\alpha\beta q_1q_2}F_5(p^2,q_1^2,q_2^2).
\end{eqnarray}
To one-loop accuracy, this form factor satisfies the spectral
representation (see, e.g., \cite{m})
\begin{eqnarray}
F_5(p^2,q_1^2,q_2^2)&=&\frac{1}{\pi}\int\limits_{4m^2}^\infty\frac{{\rm
d}s}{s-p^2-{\rm i}0}\,\Delta_5(s,q_1^2,q_2^2),\nonumber\\
\Delta_5(s,q_1^2,q_2^2)&=&\frac{N_ce_c^2m}{2\pi}\,
\frac{1}{\lambda^{1/2}(s,q_1^2,q_2^2)}
\log\!\left(\frac{s-q_1^2-q_2^2+\lambda^{1/2}(s,q_1^2,q_2^2)\sqrt{1-4m^2/s}}
{s-q_1^2-q_2^2-\lambda^{1/2}(s,q_1^2,q_2^2)\sqrt{1-4m^2/s}}\right),
\end{eqnarray}
where $\lambda\equiv(s-q_1^2-q_2^2)^2-4q_1^2q_2^2.$ The two-loop
radiative corrections to $\Delta_5(s,q_1^2,q_2^2)$ have been
calculated for massive quarks and one virtual and one real photon
and have been found to vanish \cite{teryaev2006}.

We now perform the usual steps of the method of QCD sum rules
\cite{svz}: calculate $T_{\alpha\beta}(p|q_1,q_2)$ by inserting
hadronic intermediate states, perform the Borel transform
($p^2\to\tau$), implement duality as a low-energy cut on the
corresponding Borelized spectral representation \cite{svz}, and go
to the LD limit by setting $\tau=0$ \cite{ld}. This brings us to
the representation for the $P\gamma\gamma$ form factor
\begin{eqnarray}
\label{LD_PVV}F_{P\gamma\gamma}(q_1^2,q_2^2)=\frac{2m}{M^2 f_P}
\int\limits_{4m^2}^{s_{\rm eff}}\frac{{\rm
d}s}{\pi}\,\Delta_5(s,q_1^2,q_2^2).
\end{eqnarray}
In order to obtain the form factor, we have to fix $s_{\rm eff}.$
Finding reliable criteria for fixing effective thresholds is a
rather subtle and difficult problem that has been investigated in
great detail in \cite{lms1}.

In general, the effective threshold depends on all external
kinematical variables, in our case $q_1^2$ and $q_2^2.$ We
consider both momenta as space-like and different from each other,
$q_2^2=-Q^2$ and $q_1^2=-\beta Q^2;$ therefore, we have $s_{\rm
eff}(\beta,Q^2).$ At large $Q^2$ and fixed $\beta$ the effective
threshold can be determined by matching the LD expression
(\ref{LD_PVV}) to the factorization theorem for the form factor
(\ref{factorization2}). The way how to proceed at smaller $Q^2$
will be discussed in Sec.~\ref{3.3}.

\begin{figure}[ht!]
\begin{center}
\begin{tabular}{cc}
\includegraphics[scale=.562]{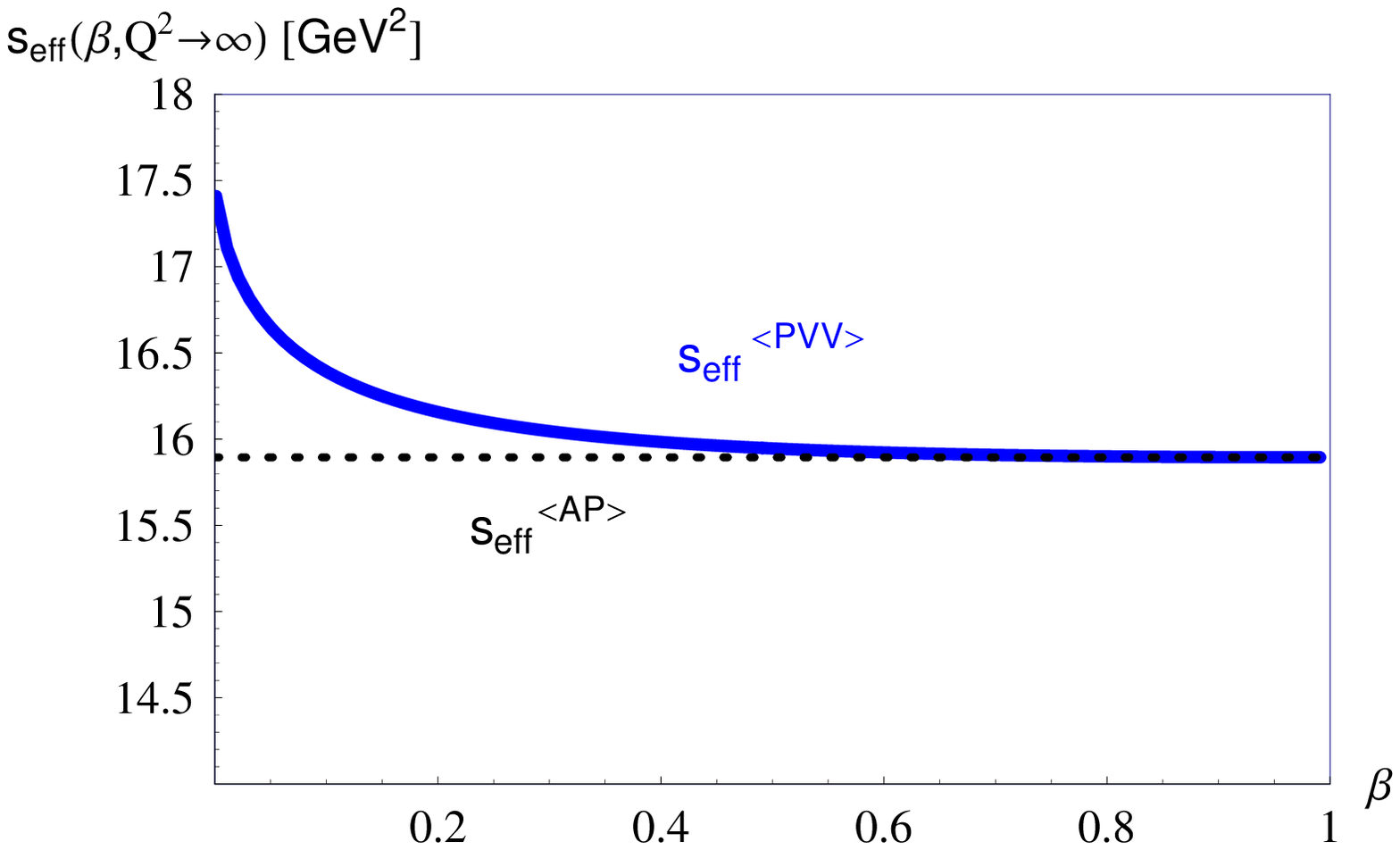}&
\includegraphics[scale=.562]{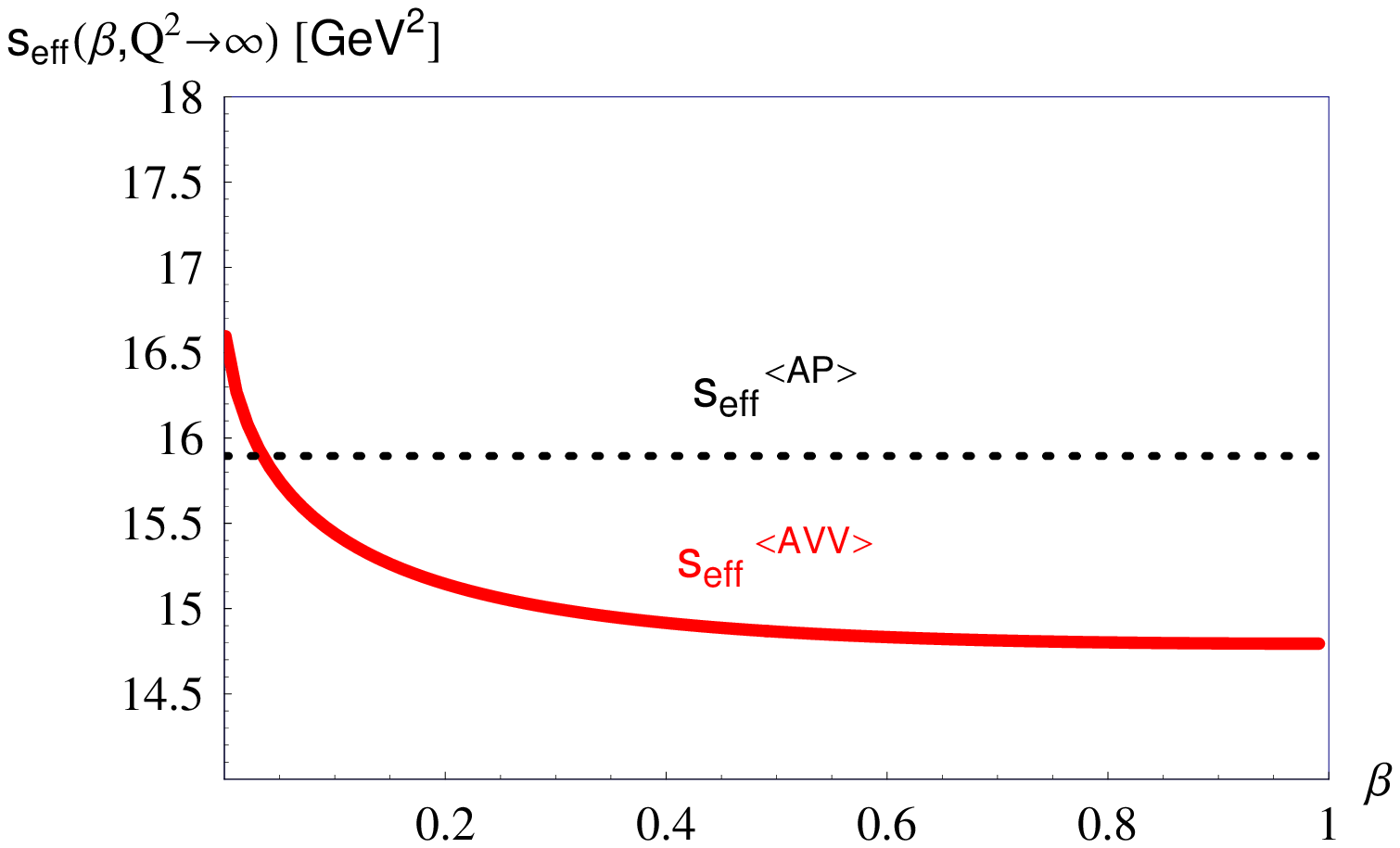}\\(a)&(b)
\end{tabular}
\caption{\label{Plot:1}Exact effective threshold $s_{\rm
eff}(\beta)\equiv s_{\rm eff}(\beta,Q^2\to\infty),$ $Q^2=Q_2^2,$
$\beta\equiv Q_1^2/Q_2^2,$ calculated by matching the LD form
factor at large $Q^2$ to the asymptotic pQCD form factor, for the
LD $\gamma^*\,\gamma^*\to P$ form factor arising from the 3-point
correlation~functions for (a) $\langle PVV\rangle$ and (b)
$\langle AVV\rangle$ (solid lines). The effective threshold for
the $\langle AP\rangle$ correlation function is indicated by
the~dashed~line.}
\end{center}
\end{figure}

For any finite effective threshold $s_{\rm eff},$ the form factor
behaves like $1/Q^2$ as demanded by pQCD. However, the spectral
density of the 3-point function $\Delta_5(s,q_1^2,q_2^2)$ does not
reduce to the product $I(\beta)\rho(s)/Q^2$ with a
$\beta$-independent function $\rho(s).$ This means that, in order
to reproduce correctly the pQCD asymptotics, the effective
threshold should depend~on $\beta.$ The result of a numerical
computation of the exact effective threshold that provides the
correct matching of the LD form factor at $Q^2\to\infty$ to the
asymptotic pQCD form factor (\ref{factorization2}) is shown in
Fig.~\ref{Plot:1}. In practice, the $\beta$-dependence~of the
threshold is not very strong.

Let us present the explicit behaviour of the
$\gamma^*\,\gamma^*\to P$ form factor for the two boundary values
$\beta=1$ and $\beta=0$~of~$\beta$:\begin{enumerate}\item For
$\beta=1,$ $Q_1^2=Q_2^2=Q^2,$ and $Q^2\to\infty,$ we find
\begin{eqnarray}
Q^2F_{P\gamma\gamma}(Q^2,Q^2)\xrightarrow[Q^2\to\infty]{}
\frac{2e_c^2}{f_P}\,\frac{N_c}{8\pi^2}\left(\frac{2m}{M}\right)^{\!2}
\int\limits_{4m^2}^{s_{\rm eff}(1,Q^2\to\infty)}{\rm
d}s\sqrt{1-\frac{4m^2}{s}};
\end{eqnarray}
thus, the effective threshold $s_{\rm eff}(1,Q^2\to\infty)$ should
be chosen equal to $s_{\rm eff}^{\rm AP}$ of the 2-point sum rule
(\ref{PP})~for~$\langle AP\rangle.$\item For $\beta=0,$ $Q_1^2=0,$
and $Q_2^2=Q^2\to\infty,$ we get
\begin{eqnarray}
Q^2F_{P\gamma\gamma}(0,Q^2)\xrightarrow[Q^2\to\infty]{}
\frac{e_c^2}{f_P}\,\frac{N_c}{4\pi^2}\left(\frac{2m}{M}\right)^{\!2}
\int\limits_{4m^2}^{s_{\rm eff}(0,Q^2\to\infty)}{\rm
d}s\log\!\left(\frac{1+v}{1-v}\right),\qquad
v\equiv\sqrt{1-\frac{4m^2}{s}}.
\end{eqnarray}
Matching to the pQCD result requires
\begin{eqnarray}
f_P^2=\frac{N_c}{24\pi^2}\left(\frac{2m}{M}\right)^{\!2}
\int\limits_{4m^2}^{s_{\rm eff}(0,Q^2\to\infty)}{\rm
d}s\log\!\left(\frac{1+v}{1-v}\right).
\end{eqnarray}
Obviously, the effective threshold $s_{\rm eff}(0,Q^2\to\infty)$
does not coincide with any of the effective thresholds for the
various 2-point functions discussed in
Sec.~\ref{Sec:2PF}.\end{enumerate}

\subsection{Transition form factor from the three-point function
$\langle AVV\rangle$}Next, we consider the amplitude for
two-photon production from the vacuum $|0\rangle,$ induced by the
axial-vector~current $j_\mu^5(x)=\bar q(x)\gamma_\mu\gamma_5q(x)$
of quarks $q$ of a single flavour:
\begin{equation}
\langle\gamma(q_1)\gamma(q_2)|j_\mu^5(x=0)|0\rangle=
T_{\mu\alpha\beta}(p|q_1,q_2)\,\varepsilon^\alpha_1\varepsilon^\beta_2,
\qquad p\equiv q_1+q_2.
\end{equation}
The amplitude $T_{\mu\alpha\beta}$ is obtained from the vacuum
expectation value of the $T$-product of one axial-vector and
two~vector currents and will be called the $\langle AVV\rangle$
amplitude. The structure of this amplitude compatible with gauge
invariance~is
\begin{eqnarray}
\label{AVV}
T_{\mu\alpha\beta}(p|q_1,q_2)=-p_\mu\epsilon_{\alpha\beta
q_1q_2}{\rm i}F_0+\left(q_1^2\epsilon_{\mu\alpha\beta
q_2}-q_{1\alpha}\epsilon_{\mu q_1\beta q_2}\right){\rm
i}F_1+\left(q_2^2\epsilon_{\mu\beta\alpha
q_1}-q_{2\beta}\epsilon_{\mu q_2\alpha q_1}\right){\rm i}F_2.
\end{eqnarray}
The form factor $F_0$ involves the contribution of the
pseudoscalar meson of interest; it can be cast into the
form~(cf.~\cite{teryaev1995})
\begin{eqnarray}
F_0(p^2,q_1^2,q_2^2)=\frac{1}{\pi}\int\limits_{4m^2}^\infty
\frac{{\rm d}s}{s-p^2-{\rm i}0}\,\Delta_0(s,q_1^2,q_2^2),
\end{eqnarray}
with the one-loop spectral density
\begin{align}
\label{delta0}
\Delta_0(s,q_1^2,q_2^2)&=\frac{m^2(q_1^2+q_2^2-s)}{\lambda}\,\Delta
+\frac{q_1^2q_2^2}{\lambda^2}
\left[(q_1^2-q_2^2)^2+(q_1^2+q_2^2)s-2s^2\right]\Delta\nonumber\\
&+\frac{1}{2\lambda^2}
\left\{(q_1^2-q_2^2)^2(q_1^2+q_2^2)-2\left[(q_1^2)^2-4q_1^2q_2^2+(q_2^2)^2\right]s
+(q_1^2+q_2^2)s^2\right\}\sigma.
\end{align}
Here,
\begin{eqnarray}
\lambda\equiv(s-q_1^2-q_2^2)^2-4q_1^2q_2^2,\qquad
\Delta\equiv\frac{1}{\pi\sqrt{\lambda}}
\log\!\left(\frac{s-q_1^2-q_2^2+\sqrt{\lambda}\sqrt{1-4m^2/s}}
{s-q_1^2-q_2^2-\sqrt{\lambda}\sqrt{1-4m^2/s}}\right),\qquad
\sigma\equiv\frac{1}{\pi}\sqrt{1-\frac{4m^2}{s}}.\qquad
\end{eqnarray}
Note that $\Delta$ and $\sigma$ are the spectral densities of the
triangle and 2-point loop diagrams with scalar particles of
mass~$m$ in the loop, respectively. One can check that
\begin{eqnarray}
\label{anomaly} \int\limits_{4m^2}^\infty\frac{{\rm
d}s}{\pi}\,\Delta_0(s,q_1^2,q_2^2)=-\frac{1}{2\pi^2},
\end{eqnarray}
independently of $q_1^2,$ $q_2^2,$ and $m^2;$ thus this integral
represents the axial anomaly \cite{teryaev1995}.

Performing the same steps as in the previous section, we obtain
the following LD expression for the $P\gamma\gamma$ form~factor:
\begin{eqnarray}
\label{LD_AVV}
F_{P\gamma\gamma}(q_1^2,q_2^2)=\frac{1}{f_P}\int\limits_{4m^2}^{\bar
s_{\rm eff}}\frac{{\rm d}s}{\pi}\,\Delta_0(s,q_1^2,q_2^2).
\end{eqnarray}
Thus, even for massive fermions the form factor is related to the
low-energy part --- the contribution below the relevant effective
threshold ${\bar s_{\rm eff}}$ --- of the axial-anomaly integral
\cite{teryaev1995}.

The two-loop radiative corrections to the $\langle AVV\rangle$
correlator vanish. This has been checked for arbitrary
virtualities~of both photons in the chiral limit \cite{2loop} and
for one real and one virtual photon for massive quarks
\cite{teryaev2006}. Multiloop~radiative corrections to the
spectral density $\Delta_0$ are unknown but expected to be nonzero
\cite{blm}. Nevertheless, the one-loop spectral density $\Delta_0$
of (\ref{delta0}) yields a reliable result for the invariant
amplitude $F_0$ in (\ref{AVV}) for not too small photon
virtualities. The principal uncertainty of the extracted
$P\gamma\gamma$ transition form factor arises from the
implementation of quark--hadron duality as a low-energy cut on the
spectral representation (\ref{LD_AVV}).

The effective threshold for the $\langle AVV\rangle$ correlator is
denoted by ${\bar s_{\rm eff}}(\beta,Q^2)$ and depends on
$q_2^2=-Q^2$ and $q_1^2=-\beta Q^2.$ As in the $\langle
PVV\rangle$ case, at large $Q^2$ the threshold can be fixed by
matching the LD result (\ref{LD_AVV}) to the pQCD asymptotics
(\ref{factorization2}). For massless and massive quarks slightly
different pictures arise, so we consider below these two cases
separately.

\subsubsection{Chiral limit}For massless quarks of a single
flavour, the spectral density takes in the limit $Q_2^2\equiv
Q^2\to\infty,$ $\beta=Q_1^2/Q_2^2$ kept~fixed, the following form:
\begin{eqnarray}
\Delta_0(s,Q_1^2,Q_2^2)\xrightarrow[Q^2\to\infty]{}
\frac{e_c^2N_cI(\beta)}{2\pi Q^2}.
\end{eqnarray}
Consequently,
\begin{eqnarray}
F_{P\gamma\gamma}(Q_1^2,Q_2^2)\xrightarrow[Q^2\to\infty]{}
\frac{2e_c^2N_cf_PI(\beta)}{Q^2}\,\frac{\bar s_{\rm
eff}(\beta,Q^2\to\infty)}{4\pi^2f^2_P}.
\end{eqnarray}
Therefore, choosing a $\beta$-independent threshold $\bar s_{\rm
eff}(\beta,Q^2\to\infty)=4\pi^2f_P^2$ reproduces the correct pQCD
asymptotics~of the form factor for any value of $\beta.$ Recall
that to order $\alpha_s$ this threshold coincides with the
effective threshold of~the~LD sum rule for the 2-point $\langle
AA\rangle$ function for massless quarks
\begin{eqnarray}
f_P^2=\frac{N_c}{12\pi^2}\int\limits_0^{s_{\rm eff}}{\rm
d}s\left[1+\frac{\alpha_s}{\pi}+O(\alpha_s^2)\right].
\end{eqnarray}
The LD model for the transition form factor at finite $Q^2$ arises
if we assume that, for all not too small $Q_1^2$ and~$Q_2^2,$ the
form factor may be well described by the LD expression
(\ref{LD_AVV}) with $\bar s_{\rm eff}=4\pi^2f_P^2.$ The form
factor at $Q_1^2=Q_2^2=0$ is related to the axial anomaly;
interestingly, this relation is satisfied for any $\bar s_{\rm
eff}(Q_1^2,Q_2^2)$ \cite{blm}. Thus, the LD sum rule with constant
$\bar s_{\rm eff}=4\pi^2f_P^2$ provides for {\em all\/} $Q_1^2$
and $Q_2^2$ the form factor $F_{P\gamma\gamma}(Q_1^2,Q_2^2)$
consistent with all rigorous~constraints. However, explicit
calculations show that for $Q^2\le2$--$4\;\mbox{GeV}^2$ the exact
effective threshold differs from its LD
value~\cite{blm}.\footnote{Setting the effective threshold equal
to $4\pi^2f_\pi^2$ in the LD sum rule for the elastic pion form
factor leads to the correct pQCD asymptotics~of $F_\pi(Q^2)$ for
$Q^2\to\infty.$ The $F_\pi(Q^2)$ data for low $Q^2,$ however,
indicate that the exact threshold at small $Q^2$ deviates from its
LD value~\cite{blm}.}

\subsubsection{Massive quarks}In this case, quark-mass corrections
destroy the nice picture one has in the chiral limit: Requiring
that, for large $Q^2,$ the LD expression reproduces the correct
pQCD asymptotics yields a $\beta$-dependent effective threshold
$\bar s_{\rm eff}$ which differs from the effective thresholds of
the 2-point correlators in Sec.~\ref{Sec:2PF}. Figure~\ref{Plot:1}
presents the exact threshold $\bar s_{\rm
eff}(\beta,Q^2\to\infty).$

The resulting explicit expressions for the two boundary values
$\beta=1$ and $\beta=0$ of $\beta$ are given below:
\begin{enumerate}\item For $\beta=1,$ one finds
\begin{eqnarray}
F_{P\gamma\gamma}(Q^2,Q^2)\xrightarrow[Q^2\to\infty]{}
\frac{4e_c^2N_cf_P}{6Q^2}\,\frac{1}{4\pi^2f_P^2}
\int\limits_{4m^2}^{\bar s_{\rm eff}(1,Q^2\to\infty)}{\rm
d}s\left(1+\frac{2m^2}{s}\right)\sqrt{1-\frac{4m^2}{s}}.
\end{eqnarray}
Therefore, in order to reproduce the correct pQCD asymptotics, we
have to require the following relation for~the effective
threshold:
\begin{eqnarray}
f_P^2=\frac{1}{4\pi^2}\int\limits_{4m^2}^{\bar s_{\rm
eff}(1,Q^2\to\infty)}{\rm
d}s\left(1+\frac{2m^2}{s}\right)\sqrt{1-\frac{4m^2}{s}}.
\end{eqnarray}
\item For $\beta=0,$ that is, for $Q_1^2=0$ and
$Q_2^2=Q^2\to\infty,$ one finds, at leading order in $1/Q^2,$
\begin{eqnarray}
F_{P\gamma\gamma}(0,Q^2)\xrightarrow[Q^2\to\infty]{}
\frac{2e_c^2N_cf_P}{Q^2}\,\frac{1}{4\pi^2f_P^2}
\int\limits_{4m^2}^{\bar s_{\rm eff}(0,Q^2\to\infty)}{\rm
d}s\sqrt{1-\frac{4m^2}{s}}.
\end{eqnarray}
Matching to pQCD requires
\begin{eqnarray}
f_P^2=\frac{1}{4\pi^2}\int\limits_{4m^2}^{\bar s_{\rm
eff}(0,Q^2\to\infty)}{\rm d}s\sqrt{1-\frac{4m^2}{s}}.
\end{eqnarray}
\end{enumerate}

\subsection{\label{3.3}Effective threshold at finite $Q^2$ and
predictions for $F_{\eta_c\gamma}(Q^2)$}Matching the LD outcomes
for the form factor to the result of the QCD factorization theorem
allows us to determine the effective thresholds at large $Q^2.$ In
order to obtain predictions for the form factor at finite $Q^2,$
we have to understand the behaviour of the effective threshold as
a function of $Q^2.$ The LD {\em model\/} for the form factor for
all $Q^2$ is obtained~by assuming that, for all not too small
$Q^2,$ $s_{\rm eff}(\beta,Q^2)=s_{\rm eff}(\beta,Q^2\to\infty).$

For the case of massless quarks, the above assumption appears
rather natural since the effective threshold found by matching LD
to pQCD at large $Q^2$ does not depend on $\beta.$ For massive
quarks, the effective threshold at large~$Q^2$~turns out to be
$\beta$-dependent. Therefore, it may not seem obvious that the
assumption of a $\beta$-dependent but $Q^2$-independent effective
threshold provides a good approximation to the exact effective
threshold.

We have tested this assumption in the case of a nonrelativistic
quantum-mechanical potential model since there the exact form
factor may be computed by solving the Schr\"odinger equation and
thus the exact effective threshold may be calculated. For the
$\bar cc$ pseudoscalar, the exact threshold at a fixed value of
$\beta$ is found to be practically $Q^2$-independent in the region
$Q^2\ge10$--$15\;\mbox{GeV}^2.$ We therefore believe that in this
region the assumption of a $Q^2$-independent threshold leads to
trustable results.\footnote{A similar analysis \cite{blm} showed
that, for light pseudoscalar mesons, the assumption of a
$Q^2$-independent threshold yields reliable results for the form
factor in the region $Q^2$ larger than a few GeV$^2.$ For smaller
$Q^2,$ the threshold may differ sizeably from the
asymptotic~threshold.}

\begin{figure}[ht!]
\begin{center}
\begin{tabular}{cc}
\includegraphics[scale=.579]{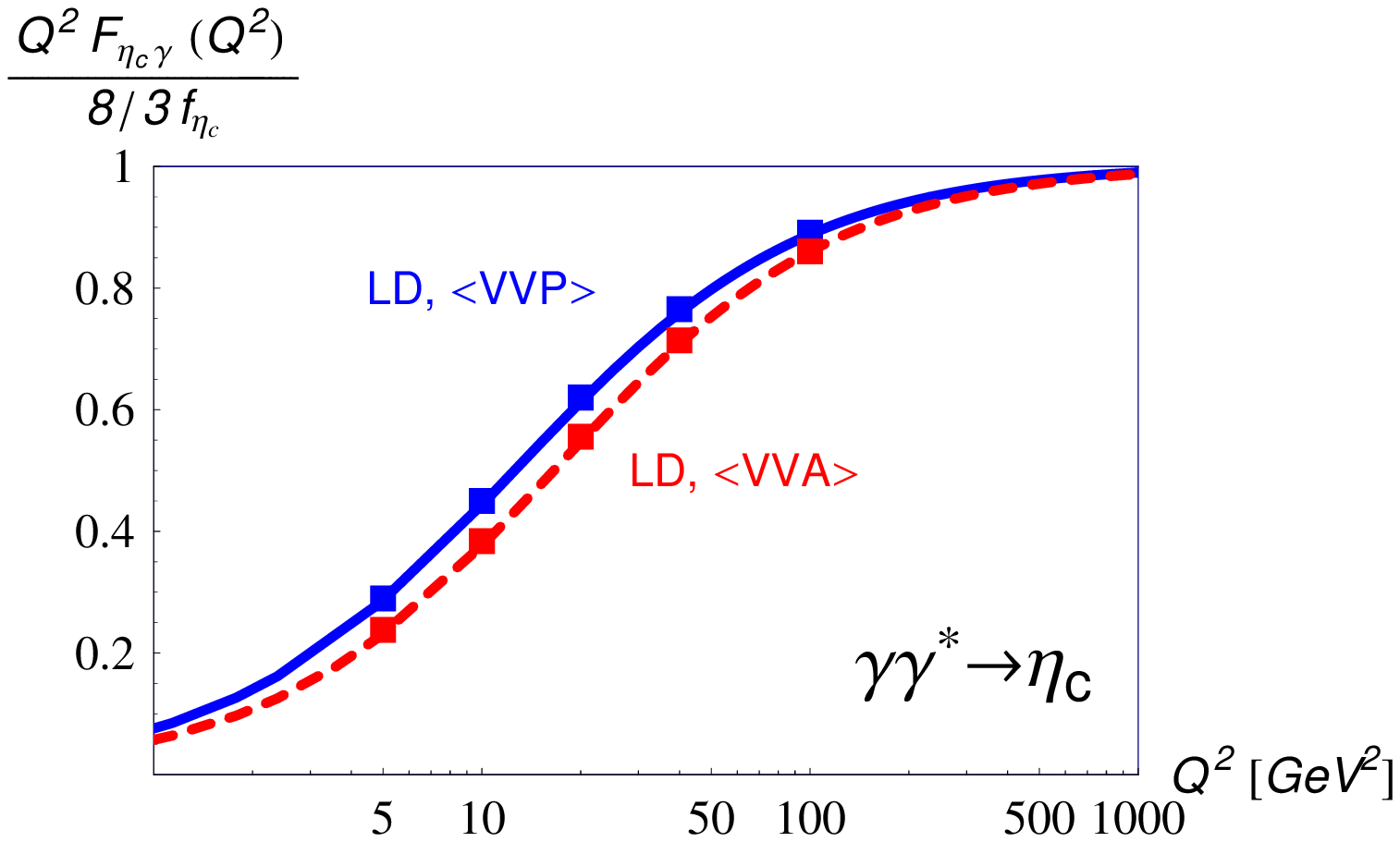}&
\includegraphics[scale=.579]{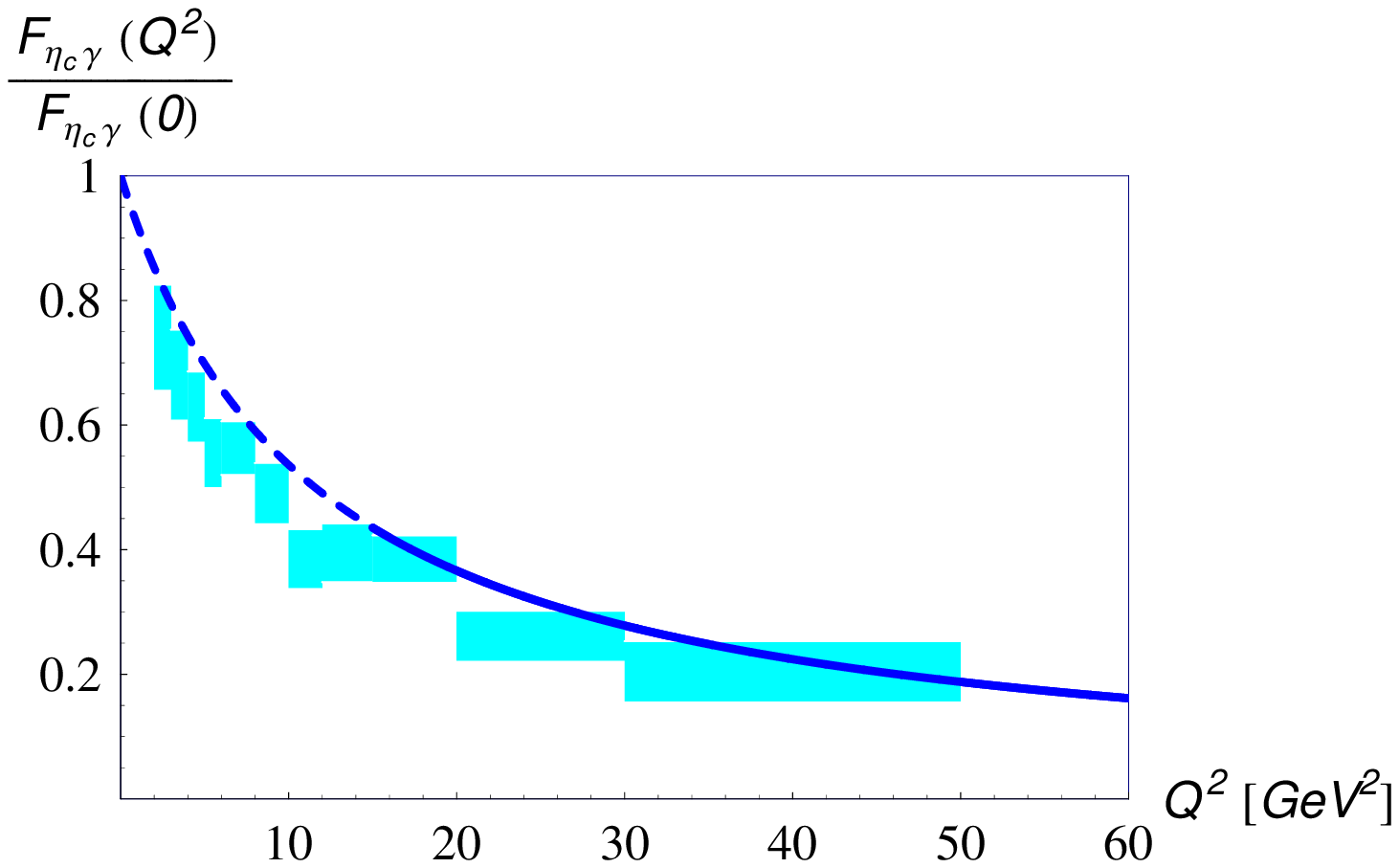}\\(a)&(b)
\end{tabular}
\caption{\label{Plot:2}Form factor $F_{\eta_c\gamma}(Q^2)\equiv
F_{\eta_c\gamma\gamma}(0,Q^2)$ for the transition
$\gamma\,\gamma^*\to\eta_c$: (a) Predictions from LD sum rules for
the correlators $\langle PVV\rangle$ (blue solid line) and
$\langle AVV\rangle$ (red dashed line). Blue and red boxes show a
monopole fit
$F_{\eta_c\gamma}(Q^2)=F_{\eta_c\gamma}(0)/(1+Q^2/M_V^2)$ to the
predictions of the LD model, with
$F_{\eta_c\gamma}(0)=8f_{\eta_c}/(3M_V^2),$ $M_V=3.5\;\mbox{GeV}$
for $\langle PVV\rangle,$ and $M_V=4.0\;\mbox{GeV}$ for $\langle
AVV\rangle.$ (b) Comparison of the LD form factor predicted by the
LD sum rule for the $\langle PVV\rangle$ correlator with recent
{\sc BaBar} measurements~\cite{babar2010}.}
\end{center}
\end{figure}

In order to get numerical estimates for the form factor, we adopt
the charm-quark mass $\overline m_c(\overline
m_c)=1.29^{+0.05}_{-0.11}\;\mbox{GeV}$~\cite{pdg} and the value
$f_{\eta_c}=0.3947\pm0.0024\;\mbox{GeV}$ of the $\eta_c$ decay
constant from lattice QCD \cite{fetac}. The corresponding
predictions for the form factors obtained from two different
correlators are shown in Fig.~\ref{Plot:2}(a). The assumption of
$Q^2$-independent thresholds for $\langle PVV\rangle$ and $\langle
AVV\rangle$ leads to a spread of predictions for
$F_{\eta_c\gamma}(Q^2)$ at finite $Q^2.$ Conservatively, this may
be regarded as an indication of the expected accuracy of the LD
model at the level of around 10\%.

We stress once more that we cannot guarantee the applicability of
the LD model at $Q^2\le10$--$15\;\mbox{GeV}^2.$ Nevertheless, let
us compare our LD predictions for small $Q^2$ with experiment.
Using $\Gamma(\eta_c\to\gamma\gamma)=7.20\pm2.12\;\mbox{keV}$
\cite{pdg}, one obtains
$F_{\eta_c\gamma}(Q^2=0)=0.08\pm0.01\;\mbox{GeV}^{-1}.$ The LD
model using the $\langle AVV\rangle$ correlator yields
$F_{\eta_c\gamma}(0)=0.067\;\mbox{GeV}^{-1};$~the LD form factor
from the $\langle PVV\rangle$ correlator has
$F_{\eta_c\gamma}(0)=0.086\;\mbox{GeV}^{-1}.$ The latter value
agrees very well with experiment, so optimistically one may expect
the LD model for $\langle PVV\rangle$ to provide reliable
predictions for the form factor for all~$Q^2.$ Figure
\ref{Plot:2}(b) compares these $\langle PVV\rangle$ results with
the available experimental data from {\sc BaBar} \cite{babar2010}.

\section{Conclusions}\label{Sec:SAC}We analyzed QCD sum rules for
the $\gamma^*\,\gamma^*\to P$ transition form factor
$F_{P\gamma\gamma}(Q_1^2,Q_2^2),$ utilizing two different 3-point
functions, $\langle AVV\rangle$ and $\langle PVV\rangle,$ in the
LD limit. We also revisited the decay constants $f_P$ of massive
$\bar qq$ pseudoscalar ground states from LD sum rules for the
2-point functions $\langle AA\rangle,$ $\langle AP\rangle,$ and
$\langle PP\rangle,$ since $f_P$ determines the asymptotics of the
form factor $F_{P\gamma\gamma}(Q_1^2,Q_2^2)$ within the framework
of pQCD factorization theorems. Our results are the following:
\begin{enumerate}\item In the LD limit, the sum rules for the
2-point functions $\langle AA\rangle,$ $\langle AP\rangle$ and
$\langle PP\rangle$ require different effective~thresholds for the
ground-state pseudoscalar meson. The sum rules and their
thresholds coincide only in the nonrelativistic limit, {\em i.e.},
for infinitely heavy quarkonia of finite radius.
\item Analyzing the form factors $F_{P\gamma\gamma}(Q_1^2,Q_2^2)$
obtained from LD QCD sum rules for the $\langle PVV\rangle$ and
$\langle AVV\rangle$ 3-point functions, we have determined the
corresponding exact effective thresholds at large momentum
transfer $Q^2=Q_2^2$ and a fixed ratio $\beta\equiv Q_1^2/Q_2^2$
by matching the LD form factors to their pQCD asymptotic behaviour
for large~$Q^2.$ These exact thresholds corresponding to
$Q^2\to\infty$ do depend on the ratio $\beta.$ This perfectly
confirms our previous findings that the effective thresholds in
QCD sum rules depend, in general, on the external kinematical
variables of the problem under consideration \cite{lms2}.
\begin{enumerate}\item The chiral limit forms the sole exception:
There the exact effective threshold for the $\langle AVV\rangle$
correlator~does not depend on $\beta$ and is equal to $s_{\rm
eff}=4\pi^2 f_P^2.$ Moreover, this effective threshold coincides
with the effective threshold of the transverse part of $\langle
AA\rangle.$\item For massive quarks, the $\beta$-dependent
effective thresholds for the 3-point functions $\langle
PVV\rangle$ and $\langle AVV\rangle$ turn out to differ from each
other and from the thresholds of the 2-point functions. Our
results for the thresholds for these two 3-point functions are
given in Fig.~\ref{Plot:1}.\end{enumerate}
\item The {\em LD model\/} for the form factor emerges if one {\em
assumes\/} that the effective threshold $s_{\rm eff}(\beta,Q^2)$
at finite~$Q^2$~does not differ sizeably from its asymptotic
behaviour $s_{\rm eff}(\beta,Q^2\to\infty).$ For light
pseudoscalar mesons, this conjecture is found to be justified for
$Q^2$ larger than a few $\mbox{GeV}^2,$ according to the results
from quantum-mechanical potential models and to the experimental
data on the $\pi\gamma,$ $\eta\gamma,$ and $\eta'\gamma$ form
factors.

For $\eta_c,$ the nonrelativistic quantum-mechanical potential
model reveals the exact effective threshold $s_{\rm
eff}(\beta,Q^2)$ to be close to $s_{\rm eff}(\beta,Q^2\to\infty)$
for $Q^2\ge10$--$15\;\mbox{GeV}^2.$ Also in QCD, the LD approach
is expected to yield reliable predictions for the $\eta_c\gamma$
transition form factor in this $Q^2$ region. Taking into account
the results for the~$F_{\eta_c\gamma}$~form factor derived from
the $\langle PVV\rangle$ and $\langle AVV\rangle$ correlators, we
conservatively estimate the accuracy~of~our predictions in this
region of $Q^2$ to be around 10\%; the accuracy improves rather
fast with rising $Q^2.$ The numerical results~for
$F_{\eta_c\gamma}(Q^2)$ from QCD sum rules may be well described
by a monopole parametrization. Combining the results~from the
$\langle PVV\rangle$ and $\langle AVV\rangle$ correlators, we
obtain
$$F_{\eta_c\gamma}(Q^2)=\frac{F_{\eta_c\gamma}(0)}{1+Q^2/M_V^2},
\qquad F_{\eta_c\gamma}(0)=\frac{2e_c^2N_cf_{\eta_c}}{M_V^2},
\qquad M_V=3.75\pm0.25\;\mbox{GeV}.$$

The Lepage--Brodsky approximate formula for the $\pi\gamma$ form
factor \cite{bl}, interpolating between the axial anomaly~at
$Q^2=0$ and the pQCD asymptotics at $Q^2\to\infty,$ too may be
cast into this form, with $M_V=2\pi f_\pi$ and the~relevant charge
factor $(e_u^2-e_d^2)/\sqrt{2}$ replacing $e_c^2.$ For the pion,
$M_V=2\pi f_\pi=0.81\;\mbox{GeV}$ is close to the $\rho$-meson
mass. Thus, the predictions of LD QCD sum rules for both light and
heavy pseudoscalars may be reasonably interpolated~by the monopole
formula
\begin{eqnarray}
F_{P\gamma}(Q^2)=\frac{2e_P^2N_cf_P}{Q^2+M_V^2},
\end{eqnarray}
with the mass parameter $M_V$ not far from the mass of the
ground-state vector meson with the relevant quantum numbers and
$e_P^2$ the corresponding charge factor.\item We investigated the
onset of the pQCD behaviour of $F_{\eta_c\gamma}(Q^2)$ and found
that, at $Q^2=100\;\mbox{GeV}^2,$ the form~factor already reaches
about 90\% of its pQCD factorization value. This conclusion does
not depend on the choice of~the correlator and is thus a solid
prediction of the LD QCD sum rules. The onset of the pQCD
behaviour of~$F_{\eta_c\gamma}(Q^2)$ is delayed with $Q^2,$
compared to the case of the light pseudoscalars. Note, however,
that we predict a much~faster onset of the pQCD regime than a
recent analysis \cite{kroll}, where the form factor at
$Q^2=100\;\mbox{GeV}^2$ reaches only~65\% of its asymptotic value.
\end{enumerate}

\vspace{3ex}{\it Acknowledgments.} We are grateful to I.~Balakireva, H.~Sazdjian, S.~Simula
and B.~Stech for valuable discussions. D.~M.\ was supported by the
Austrian Science Fund (FWF) under Project No.~P22843.


\begin{thebibliography}{30}
\bibitem{cello}
CELLO Collaboration, H.~J.~Behrend {\em et al.}, Z.~Phys.~C {\bf
49}, 401 (1991).
\bibitem{cleo}
CLEO Collaboration, J.~Gronberg {\em et al.}, Phys.~Rev.~D {\bf
57}, 33 (1998).
\bibitem{babar2006}
{\sc BaBar} Collaboration, B.~Aubert {\em et al.}, Phys.~Rev.~D
{\bf 74}, 012002 (2006).
\bibitem{babar2009}
{\sc BaBar} Collaboration, B.~Aubert {\em et al.}, Phys.~Rev.~D
{\bf 80}, 052002 (2009).
\bibitem{babar2010}
{\sc BaBar} Collaboration, J.~P.~Lees {\em et al.}, Phys.~Rev.~D
{\bf 81}, 052010 (2010).
\bibitem{babar2011}
{\sc BaBar} Collaboration, P.~del Amo Sanchez {\em et al.},
Phys.~Rev.~D {\bf 84}, 052001 (2011).
\bibitem{belle2012}
Belle Collaboration, S.~Uehara {\em et al.}, arXiv:1205.3249 [hep-ex].
\bibitem{roberts}
H.~L.~L.~Roberts, C.~D.~Roberts, A.~Bashir,
L.~X.~Guti\'errez-Guerrero, and P.~C.~Tandy, Phys.~Rev.~C {\bf
82}, 065202 (2010).
\bibitem{dorokhov}
A.~Dorokhov, JETP Lett.~{\bf 91}, 163 (2010).
\bibitem{agaev}
S.~S.~Agaev, V.~M.~Braun, N.~Offen, and F.~A.~Porkert,
Phys.~Rev.~D {\bf 83}, 054020 (2011).
\bibitem{teryaev2}
Y.~N.~Klopot, A.~G.~Oganesian, and O.~V.~Teryaev, Phys.~Lett.~B
{\bf 695}, 130 (2011); Phys.~Rev.~D {\bf 84}, 051901(R) (2011).
\bibitem{bt}
S.~J.~Brodsky, F.-G.~Cao, and G.~F.~de T\'eramond, Phys.~Rev.~D
{\bf 84}, 033001 (2011); Phys.~Rev.~D {\bf 84}, 075012 (2011);
G.~F.\ de~T\'eramond and S.~J.~Brodsky, arXiv:1203.4025 [hep-ph].
\bibitem{kroll}
P.~Kroll, Eur.~Phys.~J.~C {\bf 71} 1623 (2011).
\bibitem{mikhailov}
A.~P.~Bakulev, S.~V.~Mikhailov, A.~V.~Pimikov, and N.~G.~Stefanis,
Phys.~Rev.~D {\bf 84}, 034014 (2011); arXiv:1202.1781~[hep-ph].
\bibitem{blm}
I.~Balakireva, W.~Lucha, and D.~Melikhov, Phys.~Rev.~D {\bf 85},
036006 (2012); J.~Phys.~G {\bf 39}, 055007 (2012); W.~Lucha and
D.~Melikhov, J.~Phys.~G {\bf 39}, 045003 (2012).
\bibitem{lcqm}
C.-C.~Lih and C.-Q.~Geng, Phys.~Rev.~C {\bf 85}, 018201 (2012).
\bibitem{bl}
G.~P.~Lepage and S.~J.~Brodsky, Phys.~Rev.~D {\bf 22}, 2157
(1980).
\bibitem{anisovich}
V.~V.~Anisovich, D.~I.~Melikhov, and V.~A.~Nikonov, Phys.~Rev.~D {\bf 55}, 2918 (1997).
\bibitem{feldmann}
T.~Feldmann, P.~Kroll, and B.~Stech, Phys.~Rev.~D {\bf 58}, 114006 (1998); Phys.~Lett.~B {\bf 449}, 339 (1999).
\bibitem{ms2012}
D.~Melikhov and B.~Stech, Phys.~Rev.~D {\bf 85}, 051901 (2012).
\bibitem{ld}
V.~A.~Nesterenko and A.~V.~Radyushkin, Phys.~Lett.~B {\bf 115},
410 (1982); A.~V.~Radyushkin, Acta Phys.~Pol.~B {\bf 26}, 2067
(1995).
\bibitem{svz}
M.~A.~Shifman, A.~I.~Vainshtein, and V.~I.~Zakharov, Nucl.~Phys.~B {\bf 147}, 385 (1979).
\bibitem{m}
D.~Melikhov, Phys.~Lett.~B {\bf 380}, 363 (1996); Eur.~Phys.~J.~direct C {\bf 4}, 2 (2002) [arXiv:hep-ph/0110087].
\bibitem{lm}
W.~Lucha and D.~Melikhov, Phys.~Rev.~D {\bf 73}, 054009 (2006); Phys.~Atom.~Nucl.~{\bf 70}, 891 (2007).
\bibitem{teryaev2006}
R.~S.~Pasechnik and O.~V.~Teryaev, Phys.~Rev.~D {\bf 73}, 034017
(2006).
\bibitem{lms1}
W.~Lucha, D.~Melikhov, and S.~Simula, Phys.~Rev.~D {\bf 76},
036002 (2007); Phys.~Lett.~B {\bf 657}, 148 (2007);
Phys.~Atom.~Nucl.\ {\bf 71}, 1461 (2008); Phys.~Lett.~B {\bf 671},
445 (2009); D.~Melikhov, Phys.~Lett.~B {\bf 671}, 450 (2009).
\bibitem{teryaev1995}
O.~L.~Veretin and O.~V.~Teryaev, Phys.~Atom.~Nucl.~{\bf 58}, 2150
(1995); J.~Ho\v rej\v si and O.~Teryaev, Z.~Phys.~C {\bf 65}, 691
(1995).
\bibitem{2loop}
F.~Jegerlehner and O.~V.~Tarasov, Phys.~Lett.~B {\bf 639}, 299
(2006).
\bibitem{pdg}
K.~Nakamura {\em et al.} (Particle Data Group), J.~Phys.~G {\bf
37}, 075021 (2010).
\bibitem{fetac}
C.~H.~T.~Davies {\em et al.}, Phys.~Rev.~D {\bf 82}, 114504
(2010).
\bibitem{lms2}
W.~Lucha, D.~Melikhov, and S.~Simula, Phys.~Rev.~D {\bf 79},
096011 (2009); J.~Phys.~G {\bf 37}, 035003 (2010); Phys.~Lett.~B
{\bf 687}, 48 (2010); Phys.~Atom.~Nucl.~{\bf 73}, 1770 (2010);
J.~Phys.~G {\bf 38}, 105002 (2011); Phys.~Lett.~B {\bf 701}, 82
(2011); W.~Lucha, D.~Melikhov, H.~Sazdjian, and S.~Simula,
Phys.~Rev.~D {\bf 80}, 114028 (2009).
\end{thebibliography}
\end{document}